# Towards Sustainable DevOps: A Decision Making Framework

Muhammad Zohaib
Department of Computer Science, Govt. College University Faisalabad, Punjab, Pakistan
zohaib@gcuf.edu.pk

**Abstract**
In software industry, the DevOps is an increasingly adopting software development paradigm. Towards the sustainable DevOps adoption, there is a need to transform the organization´s Culture, Automation, Measurement and Sharing (CAMS) aspects concerning to core theme of continues development and operations. The software organizations face several complexities while implementing the DevOps principles. The sustainable DevOps implementation assist the software organizations to develop the quality projects with good return on investment. This evidence-based study aims to explore the guidelines of sustainable DevOps implementation, reported in literature and industry practices. Using systematic literature review and questionnaire survey, we identified the 48 guidelines for sustainable DevOps implementation. We further develop a decision-making framework aiming to assist the practitioners to consider the most significant set of guidelines on priority. The results show that out of CAMS, culture is the most important principle for sustainable DevOps implementation. Moreover, (i) enterprises should focus on building a collaborative culture with shared goals, (ii) assess your organization's readiness to utilize a microservices architecture and (iii) educate executives at your company about the benefits of DevOps to gain resource and budget support are the highest priority guidelines for sustainable DevOps implementation. We believe that this in-depth study helps the practitioners to understand the core principles and guidelines for sustainable DevOps implementation.

## 1. Introduction

Software industry is always looking for effective and flexible ways to develop quality software within limited time and cost. Recently, DevOps paradigm has gained popularity in software development process [1, 2]. DevOps provides platform for both development and operation teams to work collaboratively to develop software products. DevOps facilities cross functional shared responsibilities and trust between both types of development and operation teams [3]. DevOps substantially extends the continuous development goals of the agile movement by supporting automation of continuous integration and release processes[4, 5]. Leonardo et al. [6] defined DevOps as: a culture effort that automate organization infrastructure and the processing cycle of software development, guaranteeing the reliability of software product. DevOps offer several benefits to software organizations such as more focus on implementation and frequent release. Moreover, DevOps also automate the build, testing and deployment processes [7]. Forsgren[7] stated that automated development process assists to reduce the human effort and enable the automated deployment according to the schedule.

Likewise, it is emphasized that the automatic development environment significantly contributed towards the development and quality of software applications[8]. The sustainable DevOps execution allow software organizations to deliver frequent small releases which helps improve visualization of modules to the end-user[9]. The small and frequent deployment offers the development teams to receive appropriate suggestions from client which assists to modify overall quality of a product [10]. In-spite of several benefits associated with sustainable DevOps, software practitioners face numbers of challenges for the sustainability of DevOps practices such as "fear of change", "conceptual deficit", "blame game", and "complex and dynamic environments" [11]. Similarly, Ramtin et al. [12] stated that communication gap and heterogeneous environments are critical challenges for sustainable DevOps implementation in software industry.

Despite challenges associated with DevOps sustainability in software industry, several well-established organizations such as Etsy, IBM, Netflix, and Flickr have successfully adopted it[13]. For example, in Flickr, effective communication and collaboration in both development and operations practitioners have helped the organization to decrease the release time. The implementations of DevOps practices in different organizations revealed that sustainable DevOps implementation enhances the systems quality and delivery process [13, 14]. Erich et al. [13] pointed out that practices for sustainable DevOps are rapidly being adopted by software organizations with an aim to gain benefits with them. The importance of sustainable DevOps paradigm in real world practices motivated us to conduct comprehensive systematic research to investigate and analyses the guidelines reported in the state-of-art and practices. The objectives of this study are: (1) a systematic literature review and questionnaire survey approach to explore and verify guidelines of sustainable DevOps paradigm; (2) to prioritize the investigated guidelines using fuzzy-AHP approach; and (3) to develop a decision-making framework based on the rankings of guidelines. To reach-out the study objectives, the developed research questions are as follow:

[RQ1] What guidelines for sustainable DevOps implementation in software development organizations are reported in the literature and industry practices?
[RQ2] How the explored guidelines were prioritized using fuzzy-AHP?
[RQ3] What would be the prioritization-based framework for sustainable DevOps guidelines?

The paper is organized as: study background is reported in section 2. The used research methodologies are discussed in section 3. The results and analysis are presented in section 4. Summary of the study findings is shown in section 5. Section 6 presents the threats to validity of study findings. The conclusions and future direction of the study are summarized in section 7.

## 2. Background

Software organizations have shown interest in adoption of software development approaches with reduced development and delivery cycle. The basic intention behind the adoption of new development approaches is the rapid change in the customers' requirements and the consideration of requested change in positive manner. Agile development approaches have been adopted by software industry to address the rapid change concern in software development life cycle[15]. The idea and the success of continues delivery comes-up with a new software development strategy known as DevOps. DevOps is a new software development methodology which focuses on collaboration between Developer and Operation teams to work in an environment where they can share goals, processes, and tools [4, 9, 16-18]. In software industry, experts treat DevOps as the cultural movement that assists the development environment concerning with effective communication, control, and responsibilities[15, 19]. Various studies have reported that the collaboration, automation, and services are the key aspects of DevOps [9, 20].

Dyck et al. [21] mention that the revolution caused by the DevOps significantly contributed to enhance the level of trust among practitioners that assist to transform and change the development environment in software organizations. Furthermore, Smeds et al. [18] "highlighted that the DevOps is not only a culture change it also helps to improve the development process. Literature also reported the limitation and importance of DevOps paradigm[22-24]. According to Logica et al. [25], the main advantages of DevOps are product quality services and continues bonding. Similarly, Gupta et al. [26] mention that the DevOps supports in trust building between Dev and Ops practitioners. Moreover, they explored and ranked DevOps attributes that are critical to evaluate the readiness considering the adoption of DevOps in an organization. Furthermore Gill et al. [27] expressed that the DevOps contributed to develop the bridge between Dev and Ops teams that overcome the communication and coordination gape between practitioners. Anna et al. [28] argued that, the DevOps provides the roadmap for project management team to support better performance, understandability, integration, relationships among teams. However, there is a need of strong collaboration, trainings, skills and effective automation to adopt DevOps practices in a practical way." The organization adopting DevOps also faced several critical challenges[27]. Gill et al. [27] highlighted that the process and procedure-related challenges, cultural conflicts, and the problems in operational models.

The existing literature portrays evidence-based research in the context to explore the guidelines for DevOps sustainability in software organizations. Furthermore, no research has been done to analyze the sustainable DevOps guidelines using the fuzzy-AHP approach. This detailed empirical investigations and analysis, will help the teams to understand and develop the methodologies for sustainable implementation of DevOps in software development industry.

## 3. Research design

The research design is divided into three steps, (1) systematic literature review (SLR), (2) questionnaire survey study, and (3) fuzzy AHP. At initial phase, the SLR was undertaken with the goal of identifying the guidelines for sustainable DevOps. Second, a questionnaire poll was conducted to obtain input from industry practitioners on the identified guidelines. The fuzzy-AHP was used in the third stage to rank the guidelines based on their importance for sustainable DevOps. Figure 1 depicts the research design in detail.

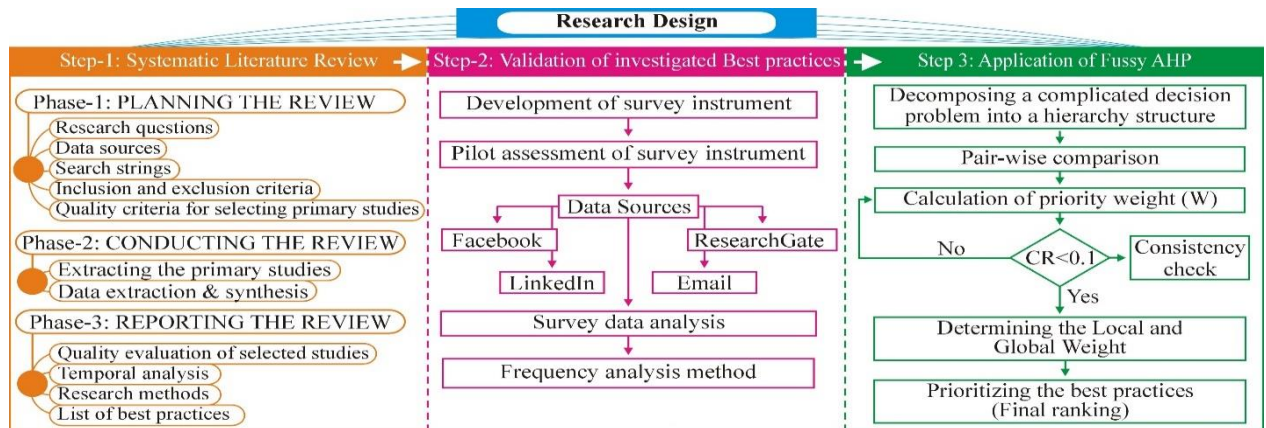
Figure 1: Study Research Design

### 3.1. Systematic Literature Review (SLR)

SLR is considered to collect the most related literature concerning the research questions and objective of the study. SLR is considered as an effective approach to collect and evaluate the potential literature related to a specific topic. The SLR procedural approach that provides more valid and comprehensive results than informal literature review[29]. In this study, we have used the procedures defined by Kitchenham and Charters[29] to perform SLR. Considering the recommendations of Kitchenham and Charters[29], the review process consists of three broad phases: "planning the review", "conducting the review" and "reporting the review". The detailed steps of the SLR approach are presented in Figure 1 and described in the following sections.

#### 3.1.1. Planning the review

Planning refers to developing the protocols adopted to collect and analyses the data. The following review protocols were adopted to extract and analyses the literature to answer the proposed research question.

**(i) Search sources**

*Data collection source:*

To collect the potential literature related to the research objective of the study, the selection of appropriate data sources is essential. Though, the suggestions of Chen et al. [30] and Zheng et al.[31] were considered, and 8 digital-databases were used to explore the data. Figure 2 presents the selected libraries along with the number of literatures comes against the execution of search string.

*Search string:*

An effective search string is essential to collect literature related to study objective. To develop the search string, we have used the key terms and their alternatives, collected from the existing studies[1, 13, 25, 27, 32], by following the guidelines of [31, 33]. Using the OR and AND operators were used to formulate the complete string, as presented below:

("guidelines" OR "practices" OR "motivators" OR "activities" OR "Concerns" OR "techniques" OR "tools" OR "methods" OR "process" OR "evaluation") AND ("DevOps" OR "Development and Operation" OR "Continues development and operation".

*Initial inclusion criteria:*

The protocols were created to determine whether to include literature gathered throughout the literature extraction process. The protocols for inclusion were taken from current studies. Inayat et al. [34] and Niazi et al. [35]. (1) "The article should be submitted to a reputable journal, conference, or book chapter for publication." (2) "The essay should discuss the obstacles that hinder DevOps implementation." (3) The findings of the study are based on empirical data sets. (4) The report should clearly state why DevOps adoption is important. (5) Selected literature must be written in English.

*Initial exclusion criteria:*

We've refined the protocols to omit the literature gathered from databases at the outset. The exclusion criteria were determined based on previous research. Inayat et al. [34], Niazi et al. [35], and Akbar et al. [36] (1) "Only the most completed study from a similar research endeavor was taken into account." (2) The article should include specific details on DevOps implementation. (3) The study that has nothing to do with the research's goal. (4) The paper used in the study is full or regular. (5) The studies from the literature review were not considered.

*Study quality assessment (QA):*

The purpose of the QA was to determine the adequacy of the chosen study for the study objective. The QA is carried out in accordance with Kitchemhm and Charctros's guidelines[29]. The five-questions QA method were developed and evaluated using the Likert scale, if the study fully answers the criteria then the assigned score is 1, for partial=0.5 and 0-score if the study does not give any information about the developed criteria. Several previous studies[34-36] have used similar criteria. Appendix-A contains the outcomes of the QA.

Table 1: QA criteria

| Sr# | Criteria |
|---|---|
| QA1 | Does the used research method in the selected literature aligned with RQs? |
| QA2 | Does the collected reported the DevOps guidelines? |
| QA3 | Does study elaborate the details of DevOps adoption? |
| QA4 | Are the reported guidelines related to DevOps project management? |
| QA5 | Are the study findings justify the research questions? |

### 3.1.2 Conducting the review
*Final study selection:*

Primarily, 860 studies were extracted in the response of the search string executed on the selected database. The collected literature was further refined by applying the phase of tollgate approach developed by Afzal et al. [37]. The tollgate approach consists of five phases, and each step is performed carefully, aiming to select the studies for data extraction finally.

Though, as presented in Figure 2, a total of 71 studies were selected for the final data extraction process. All the conclusive selected studies were assessed concerning their significance for addressing the research questions of the survey. The excellent reviews were labelled as 'SP' to present their usage in the paper. The list of selected studies and their QA score is given in Appendix-A.

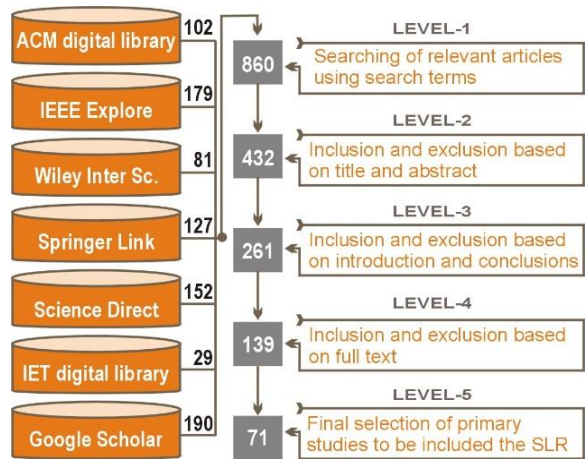

Figure 2: Final selected studies

*Data extraction and synthesis:*

Finally, 71 studies (Figure 2) were thoroughly reviewed for data extraction that corresponded to the research goal. For data extraction, author no.1 and 2 are continually participated, the third and fourth authors validated the extracted data. Initially, the selected studies provided the claims, primary themes, concepts, practices, and actions. The collected data was then synthesized into concise statements, resulting in the final 48 DevOps implementation recommendations.

Although there may be bias in the study outcomes, the "inter-rater reliability test" was used to assess the mapping team's baseness[37]. Three external specialists were requested to participate in the mapping process to help with this. They chose the 12 research at random and began the data extraction process. We developed the non-parametric "Kendall's coefficient of concordance (W) [38] based on the findings of study authors and external experts. "A W=1 number implies perfect agreement, while a W=0 value indicates entire disagreement." The results of W=0.84 p=0.003 suggest that the study authors and external experts have significant agreement significantly. This implied that the study's findings are consistent. The used code is given in this link: *https://tinyurl.com/y5fct4ql.*

### 3.1.3 Reporting the review

*Quality of selected studies:*

The QA of the selected studies demonstrates how the literature is effective in answering the study's research questions. According to the QA cumulative results, more than 70% of studies receive a score of 70% or higher. Appendix-A contains the detailed QA results. We selected a 50 percent score as a cutoff point in this study.

*List of guidelines:*

During data extraction process, the concepts, themes, and ideas reported in the selected literature were extracted, and by paraphrasing, we develop a list of 48 guidelines that are important for sustainable DevOps implementation in software development organizations.

### 3.2. Empirical study

To collect the perceptions of industry experts, we used the questionnaire survey approach, and the used steps are discussed in the sub-sequent sections:

#### 3.2.1 Questionnaire development

An online survey questionnaire was created using the Google Form platform (docs.google.com/forms) to verify the SLR findings. The questionnaire survey is divided into three sections: the first section contains questions about the survey participants' bibliographic information; the second section contains questions about the survey participants' organizational information; and the third part is closed-ended and contains the SLR-identified list of DevOps guidelines. (3) The fourth portion is open-ended, and it allows survey respondents to add any extra guidelines that was not included in the closed-ended section. The participants feedback was received in the form of five-point Likert scale, from Strongly agree to strongly disagree. The Likert-scale also includes neutral option which help the practitioners to give unbiased feedback, as without neutral option, the respondents are banned to go to agree or disagree side[39].

#### 3.2.2 Pilot testing

The pilot assessment of the questionnaire was conducted to check and improve the understandability of questionnaire questions [40-42]. To do this, the developed questionnaire was sent to three experts, one experts from academic (Chongqing University, Chain) and two belongs to industry practices ("Virtual force-Pakistan" and "QSoft-Vietnam"). The experts recommend several changes to the questionnaire layout and questions for collecting bibliographic data from survey respondents. They also recommend arranging the questions in a table format. We updated questionnaire according to experts' opinions and the updated one is used for data collection process. The sample of used questionnaire is given in Appendix-B.

#### 3.2.3 Data sources

The data sources are crucial in determining the target population. The potential community must be targeted because it is required for the collection of accurate data. The goal of this survey study was to gain expert opinion into DevOps difficult factors discovered through literature using SLR study. We used professional email addresses, Research-Gate, and LinkedIn to target the population. The survey questionnaire was distributed to the intended geographically dispersed group using the snowballing technique [40-42].

From December 2020 to May 2021, the data collection process was carried out. A total of 102 replies were collected during the data gathering process. All the responses were manually verified, and nine were found to be incomplete. We opted not to include the incomplete response in the data analysis process after consulting with the study team. The final 93 complete replies, on the other hand, were employed in the data analysis procedure. The bibliographic information is presented in section-4.2.

#### 3.2.5 Survey data analysis

Frequency data analysis approach was adopted to analyze the collected responses, as it is considered the effective way to compares the responders opinions in between the variables and across the group of variables [43]. The same approach has been adopted in the existing studies [44-46].

### 3.3. Phase 3: Fuzzy Set Theory and AHP

#### 3.3.1 Fuzzy set theory

The Fuzzy set theory is an extended version of classical set theory that's initially proposed by Zadeh[47]. That was oriented to fix the vagueness of uncertainties of ear world practices using multicriteria decision making problems.

The basic input of Fuzzy set theory is to epitomize the vague data, where's the membership function µF(x) objects maps between 0 and 1. The protocols of fuzzy set theory along with definition are presented in sub-sequent sections:

Definition: "A triangular fuzzy number (TFN) F is denoted by a set" (vl, vm, vu), as shown in Figure 4 and the membership-function donated as µF(x) of F.

$$\mu_F(x) = \begin{cases} \dfrac{t - v^l}{v^m - v^l}, & v^l \leq t \leq v^m \\ \dfrac{v^u - t}{v^u - v^m}, & v^m \leq t \leq v^u \\ 0, & Otherwise \end{cases} \quad (1)$$

Where $v^l$, $v^m$ and $v^u$ denotes the crisp lowest, highest priority, and highest possible values.

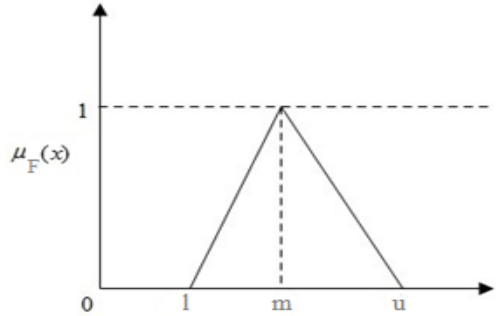

**Figure 4.** Triangular fuzzy number

The algebraic operational laws using two TFNs, namely ($V_1$, $V_2$) are given in Table 2.

Table 2. Triangular Fuzzy Numbers

| Operation Law | Expression |
|---|---|
| Addition ($V_1 \oplus V_2$) | $(v^l_1, v^m_1, v^u_1) \oplus (v^l_2, v^m_2, v^u_2) = (v^l_1 + v^l_2, v^m_1 + v^m_2, v^u_1 + v^u_2)$ |
| Subtraction ($V_1 \oplus V_2$) | $(v^l_1, v^m_1, v^u_1) \oplus (v^l_2, v^m_2, v^u_2) = (v^l_1 - v^l_2, v^m_1 - v^m_2, v^u_1 - v^u_2)$ |
| Multiplication ($V_1 \oplus V_2$) | $(v^l_1, v^m_1, v^u_1) \oplus (v^l_2, v^m_2, v^u_2) = (v^l_1 * v^l_2, v^m_1 * v^m_2, v^u_1 * v^u_2)$ |
| Division ($V_1 \oplus V_2$) | $(v^l_1, v^m_1, v^u_1) \oplus (v^l_2, v^m_2, v^u_2) = (v^l_1 / v^l_2, v^m_1 / v^m_2, v^u_1 / v^u_2)$ |
| Inverse ($V_1 \oplus V_2$) | $(v^l_1, v^m_1, v^u_1)^{-1} = (1/v^l_1, 1/v^m_1, 1/v^u_1)$ |
| For any real number k ($kV_1$) | $k(v^l_1, v^m_1, v^u_1) = k\,v^l_1, k\,v^m_1, k\,v^u_1$ |

*3.3.2 Fuzzy analytical hierarchy process (FAHP)*

FAHP is the most effective and powerful approach used to fix the multicriteria decision making problems. The key benefit of FAHP is the relative ease with which it manages the multiple criteria, easier to understand, and it can efficiently manage both qualitative and qualitative data. The following primary step of FAHP approach:

**Step1:** Develop a hierarchy structure of the decision-making problem (as given Figure 5)
**Step2:** Use pairwise comparison and calculate the weights.
**Step3:** Apply the consistency check.
**Step4:** Determine the priority order of each guideline.

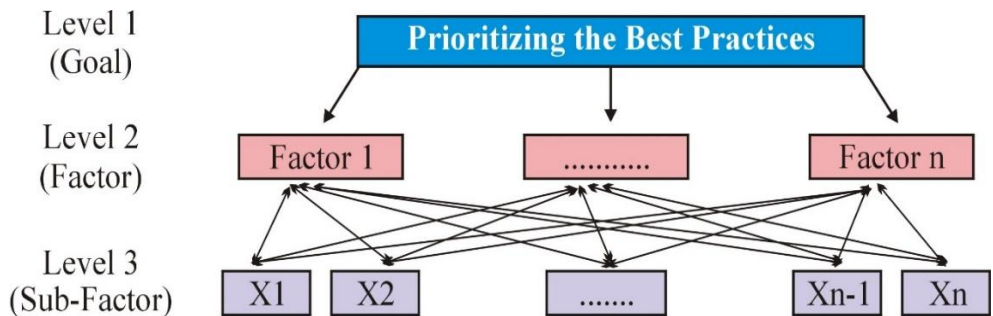

**Figure 5.** FAHP decision hierarchy

However, conventional AHP has numerous advantages[48-50], but it also faced some core limitations as it is based on the "Crisp environment", "Judgmental scale is unbalanced", and the "absence of uncertainty", and because of these limitations the selection of judgment is subjective. The FAHP was developed to address these limitations of AHP to get results more effectively and accurately [51]. The FAHP deals with uncertainties, imprecise judgment of different experts by handling the linguistic variables. FAHP approach has been considered in different context [51-54]. To address the uncertainties and vagueness we have used the FAHP suggested by Chang [55] that provides more appropriate and consistent results compared with other FAHP approaches.

In a prioritization problems, let X = $\{v_1, v_2,..., v_n\}$ signify the elements of main categories as an object set and U = $\{t_1, t_2,..., t_n\}$ presents the values of a particular category as a goal set. Considering the Chang [55] approach, every object is measured, and extent-analysis for objective (gi) is performed, respectively. The following Equation" (2) and (3) are used to generate (m) extent analysis values for each object:

$$V^1_{gi}, V^2_{gi}, \ldots, V^m_{gi}, \quad (2)$$
$$i = 1, 2, \ldots, n \quad (3)$$

Where, $F^j_{gi}$ (j = 1, 2, ..., m) are fuzzy triangular numbers (TFNs). The Chang's extent analysis [55] is performed in the following steps:

**Step 1:** The element of fuzzy synthetic extent ($S_i$) for the $i^{th}$ object using Eq. (4):

$$S_i = \sum_{j=1}^{m} V^j_{gi} \otimes \left[ \sum_{i=1}^{n} \sum_{j=1}^{m} V^j_{gi} \right]^{-1} \quad (4)$$

To achieve the expression $\sum_{j=1}^{m} V^j_{gi}$, execute the fuzzy addition operation of m extent analysis using Eq. (5):

$$\sum_{j=1}^{m} V^j_{gi} = (\sum_{j=1}^{m} v^l_{gi}, \sum_{j=1}^{m} v^m_{gi}, \sum_{j=1}^{m} v^u_{gi}) \quad (5)$$

and to make the expression $\left[ \sum_{i=1}^{n} \sum_{j=1}^{m} V^j_{gi} \right]^{-1}$, the fuzzy addition operation is performed on $V^j_{gi}(j=1,2,\ldots.m)$ value, as follow using Eq. (6):

$$\sum_{i=1}^{n} \sum_{j=1}^{m} V^j_{gi} = (\sum_{i=1}^{n} v^l_i, \sum_{i=1}^{n} v^m_i, \sum_{i=1}^{n} v^u_i) \quad (6)$$

To end, The inverse of each vector is determined using Eq. (7):

$$\left[ \sum_{i=1}^{n} \sum_{j=1}^{m} V^j_{gi} \right]^{-1} = (\frac{1}{\sum_{i=1}^{n} v^u_i}, \frac{1}{\sum_{i=1}^{n} v^m_i}, \frac{1}{\sum_{i=1}^{n} v^l_i}) \quad (7)$$

**Step 2:** As $F_a$ and $F_b$ are two fuzzy triangular numbers, then these fuzzy numbers need to be compared that is knows as Degree of possibility i.e. $V_a = (v^l_a, v^m_a, v^u_a) \geq V_b = (v^l_b, v^m_b, v^u_b)$ and is compared as follows using Eq.(8) and Eq. (9).

$$V(V_a \geq V_b) = sup[min(\mu_{va}(x), (\mu_{vb}(x))] \quad (8)$$

$$V(V_a \geq V_b) = hgt(V_a \cap V_b) = \mu_{v_a}(d) = \begin{cases} 1 & \text{if } v^m_b \geq v^m_a \\ \frac{v^l_b - v^u_a}{(v^m_b - v^u_a) + (v^m_b - v^l_b)} & \text{Otherwise} \\ 0 & v^l_a \leq v^u_b \end{cases} \quad (9)$$

Figure 6 indicates the highest intersection point in D, $\mu_{Va}$ and $\mu_{Vb}$ (Figure 6). The values of $T_1(V_a \geq V_b)$ and $T_2(V_a \geq V_b)$ are compulsory for calculating the value of $P_1$ and $P_2$.

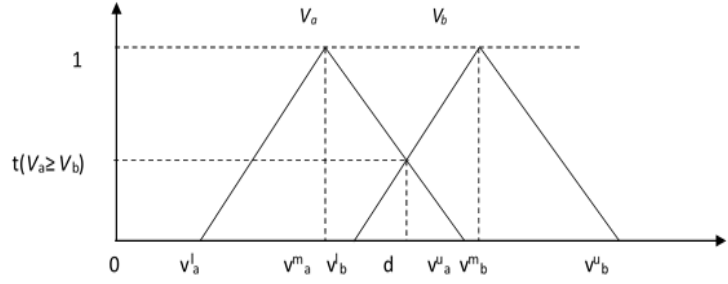
Figure 6. Triangular Fuzzy number

Table 3: RI against each matrix size

| "Size of the matrix" | 1 | 2 | 3 | 4 | 5 | 6 | 7 | 8 | 9 | 10 | 11 | 12 | 13 | 14 | 15 |
|---|---|---|---|---|---|---|---|---|---|---|---|---|---|---|---|
| "Random consistency index" (RI) | 0 | 0 | 0.58 | 0.9 | 1.12 | 1.24 | 1.32 | 1.41 | 1.45 | 1.49 | 1.51 | 1.48 | 1.56 | 1.57 | 1.59 |

**Step 3:** Calculate the probability that a convex fuzzy number is bigger than k convex fuzzy numbers $V_i$ ($i = 1, 2,..., k$) can be calculated as follow using Eq. (10) and Eq.(11):

$$T(V \geq V_1, V_2, V_3....V_k) = \min T(V \geq V_i) \quad (10)$$

Assuming that,

$$d'(V_i) = \min T(V_i \geq V_k) \quad (11)$$

for $k = 1,2,...,n; k \neq i$.

With the help of Eq. 12, calculate the weight vector.

$$W' = (d'(V_1), d'(V_2), d'(V_3),.....d'(V_n)) \quad (12)$$

Where, $V_i$ ($i = 1,2,...,n$) are $n$ distinct elements.

**Step 4:** The normalized weight vectors are calculated using Equation 13, and the result will be a non-fuzzy number (known as defuzzification) which represents priority weight for the criteria:

$$W = (d(V_1), d(V_2), d(V_3),.....d(V_n)) \quad (13)$$

Where $W$ is a non-fuzzy number.

**Step 5: Checking consistency ratio**:
In fuzzy AHP [56], the pairwise matrices should always be consistent. As a result, the consistency ratio of each pairwise comparison matrix must be checked [57, 58]. The graded mean integration technique is used to defuzzify the matrix in this way. $P = (l, m, u)$ is a triangular fuzzy number that can be defuzzified to a crisp number as follows:"

$$P_{crisp} = \frac{(4m + l + u)}{6} \quad (14)$$

After each value in the matrix has been defuzzified, the consistency ratio (CR) of the matrix may be simply calculated and tested to see if it is less than 0.10. Two key parameters are employed for this, namely the consistency index (CI) and the consistency ratio (CR), which are defined using Equations 14 and 15, respectively.

$$CI = \frac{I_{max} - n}{n - 1} \quad (15)$$

$$CR = \frac{CI}{RI} \qquad (16)$$

## 4. Results and analysis

By carefully applying the literature collection and data extraction process, a total of 48 guidelines were identified, which are reported in the literature. Bajpai [59] emphasized that for sustainable DevOps implementation, the software development organization should focused on core principle of DevOps i.e., culture, automation, measurement and sharing (CAMS). Originally, the CAMS framework was introduced by Kim[60] to define the core pillars of DevOps.

***Culture:*** Adapting to a DevOps culture entails converting into a company that prioritizes providing high-quality service. Service is not just making a product available to clients, but also ensuring that the product is in excellent functioning order and consistently meets their demands. To deliver better service, teams should understand how customers will receive and use the product. This will enable the teams to provide ongoing assistance to clients even after the product has been delivered. Developing a DevOps mindset - DevOps teams are motivated by certain traits to focus on providing high-quality service to clients. They are an environment that encourages continuous learning, experimentation, a product-oriented attitude, an engineering culture, and a quality-oriented focus.

***Automation:*** Routine tasks will be automated, resulting in a shift in which members of the Software Delivery team will primarily execute jobs with high variability and analyzability. To put it another way, engineering tasks. This necessitates a company with an organic structure and engineering teams that are mainly independent, multidisciplinary, and self-organized. To add value to the table, organizations must employ DevOps principles. Processes that are automated are repeatable. Process execution costs are kept to a minimum. To put it another way, automation produces predictable and standardized results. These ideas can be used to improve software delivery processes.

***Measurement:*** Continuously providing value and improvement requires systematic measurement process. The process can be improved by keeping track of key parameters and providing feedback. As part of any DevOps implementation, measurement and monitoring must be integrated into your day-to-day procedures. Callanan and Spillane [17] highlights the importance of bidirectional feedback loops for obtaining feedback from the development and operations teams.

***Sharing:*** DevOps is the collaboration of development and operations and sharing is fundamental to this. DevOps requires ongoing collaboration and, most crucially, knowledge exchange. Every team in an organization has experienced failures in terms of people, procedures, tools, projects, and so on. To support successful DevOps deployment, this knowledge should be shared among team members within an organization.

We mapped the investigated guidelines into CAMS principles as it covers all the important aspects of DevOps[61]. the coding scheme [62] was used to perform the mapping process. The mapping team consist of three authors of this study (Author no.1, 3 and 4), and two experts were invited from the real-world industry practices the mapping team labelled and grouped the guidelines into their most related categories. The mapping results are given in Table 4.

We conducted an "inter-rater reliability test" between the mapping team and two external experts to measure the researchers bias (invited from empirical software engineering labs). External experts were tasked with categorizing the collection of rules into the most closely related categories. We calculated the "non-parametric Kendalls coefficient of concordance" (W)[13] using the results of both the mapping team and external experts to check the inter-rater agreement between the researchers and independent experts. W=1 indicates complete agreement, whereas W=0 indicates complete disagreement. The results (W=0.89, p=0.005) demonstrated that researchers' and independent experts' mapping processes are similar. This indicates that the researcher and independent specialists have reached an agreement on the mapping method. As a result, the mapping is unbiased and consistent.

Table 4: List of identified guidelines and their mapping in CAMS

| Categories | Sr# | Guidelines | IDs of selected SLR studies |
|---|---|---|---|
| Measurement | G1 | Organizations start DevOps practices with small projects | PS1, PS3, PS13, PS 16, PS19, PS 21, PS29, PS33, PS37, PS41, PS46, PS52, PS55, PS58, PS64, PS70 |
| | G2 | Include modeling for legacy infrastructure and applications in your DevOps plans | PS6, PS15, PS 18, PS25, PS 31, PS39, PS43, PS54, PS67 |
| | G3 | Consider application architecture changes based on on-premises, cloud, and containers early on in the | PS9, PS17, PS27, PS 36, PS44, PS 55, PS70 |
| | G4 | Avoid fragmented toolset adoption, which can add to your costs | PS2, PS5, PS10, PS 12, PS17, PS 24, PS28, PS34, PS38, PS44, PS45, PS50, PS54, PS57, PS61, PS68 |
| | G5 | Effective and comprehensive measurement and monitoring | PS5, PS10, PS 14, PS18, PS 25, PS27, PS35, PS39, PS43, PS46, PS50, PS55, PS61, PS64, PS66 |
| | G6 | Decide which processes and tests to automate first | PS1, PS7, PS 17, PS26, PS 31, PS43, PS44, PS56, PS61, PS70 |
| | G7 | Monitor the Application's Performance | PS4, PS7, PS11, PS20, PS 22, PS28, PS34, PS38, PS42, PS45, PS49, PS53, PS54, PS56, PS59, PS66, PS70, PS71 |
| | G8 | Integrated Configuration Management | PS11, PS22, PS31, PS 42, PS53, PS59, PS69 |
| | G9 | Emphasize Quality Assurance Early | PS6, PS9, PS 15, PS23, PS 29, PS31, PS37, PS42, PS45, PS54, PS60, PS63, PS67 |
| | G10 | Active Stakeholder Participation | PS5, PS13, PS 19, PS27, PS 30, PS36, PS43, PS49, PS55, |
| | G11 | Use tools to capture every request | PS8, PS18, PS 25, PS33, PS 45, PS52, PS60, PS61 |
| Automation | G12 | Decide which processes and tests to automate first | PS4, PS8, PS 12, PS16, PS 26, PS30, PS36, PS38, PS40, PS45, PS52, PS56, PS62, PS68 |
| | G13 | Continuous integration and testing | PS4, PS5, PS 19, PS30, PS 41, PS47, PS52, PS65, PS70 |
| | G14 | Implement tracking and version control tools | PS 9, PS14, PS 24, PS28, PS36, PS37, PS42, PS51, PS59, PS62, PS66 |
| | G15 | Have a centralized unit for DevOps | PS5, PS8, PS13, PS 15, PS23, PS 24, PS30, PS35, PS37, PS40, PS44, PS50, PS57, PS64 |
| | G16 | Reduce handoffs | PS3, PS9, PS14, PS 16, PS21, PS 23, PS29, PS36, PS37, PS41, PS46, PS48, PS52, PS57, PS68, PS69 |
| | G17 | Implement Automation in Dashboards | PS2, PS11, PS14, PS30, PS 38, PS45, PS54, PS64, PS66 |
| | G18 | Use the right and advanced tools | PS7, PS15, PS16, PS 22, P25, PS 31, PS34, PS39, PS42, PS47, PS55, PS61, PS69 |
| | G19 | Use tools to capture every request | PS9, PS17, PS 23, PS35, PS 41, PS47, PS53, PS64 |
| | G20 | Use tools to log metrics on both manual and automated processes | PS7, PS8, PS 13, PS24, PS 27, PS32, PS39, PS43, PS46, PS51, PS60 |
| | G21 | Provisioning and change management | PS3, PS6, PS15, PS 17, PS19, PS 25, PS29, PS35, PS39, PS45, PS47, PS51, PS55, PS56, PS62, PS67, PS69, PS70 |
| | G22 | Build Up the Rest of Your CI/CD Pipeline | PS6, PS11, PS18, PS 24, PS28, PS 34, PS35, PS39, PS45, PS52, PS57, PS70 |
| | G23 | Take a 'security first approach' | PS11, PS19, PS 26, PS34, PS 42, PS51, PS59 |
| | G24 | Use on-demand testing environments | PS3, PS10, PS 17, PS20, PS 25, PS32, PS35, PS41, PS46, PS50, PS58, PS60, PS62 |
| | G25 | Develop automated continues deployment environment | PS2, PS7, PS9, PS 11, PS13, PS 16, PS23, PS26, PS29, PS34, PS37, PS41, PS45, PS52, PS58, PS64, PS68, PS70 |
| | G26 | Standardize and automate complex DevOps environments with cloud sandboxes and other tools | PS6, PS11, PS 15, PS22, PS 26, PS31, PS38, PS42, PS44, PS49, PS51, PS53, PS56,PS61, PS65 |
| Sharing | G27 | Ensure continuous feedback between the teams to spot gaps, issues, and inefficiencies | PS8, PS13, PS15, PS 23, P26, PS 29, PS35, PS38, PS45, PS49, PS56, PS69 |
| | G28 | Communications and collaboration planning | PS4, PS4, PS8, PS 12, PS15, PS 19, PS22, PS28, PS29, PS31, PS35, PS39, PS43, PS54, PS59, PS65, PS69, PS70, PS71 |
| | G29 | Continuous practice and planning to avoid resistance | PS13, PS17, PS 25, PS36, PS 45, PS53, PS62 |
| | G30 | Create real-time project visibility | PS5, PS 20, PS31, PS37, PS44, PS48, PS52, PS54, , PS66 |
| | G31 | Increase flow of communication by reducing batch size | PS1, PS7, PS 16, PS21, PS 27, PS33, PS35, PS43, PS48, PS56, PS67 |
| | G32 | Building trust and share values and goals for effective channel | PS3, PS4, PS10, PS 15, PS16, PS 20, PS24, PS27, PS30, PS32, PS36, PS38, PS45, PS46, PS52, PS54, PS62, PS65, PS66, PS68 |
| | G33 | Enterprises should standardized processes and establish common operational procedures | PS9, PS11, PS 17, PS22, PS 25, PS30, PS34, PS41, PS47, PS49, PS51 |
| | G34 | Create a clear plan that includes milestones, project owners, and well-defined deliverables | PS4, PS8, PS 14, PS22, PS 25, PS30, PS34, PS41, PS47, PS49, PS51, PS55, PS56, PS63 |
| | G35 | Teams need training on DevOps | PS5, PS12, PS 21, PS32, PS35, PS42, PS45, PS51, PS55, PS61, PS71 |
| | G36 | Shared code of conduct, a formal roles ssignment, and clear and simple processes may help in understanding responsibilities | PS1, PS9, PS 18, PS24, PS32, PS40, PS46, PS52, PS56, PS60, PS62, PS 69, PS71 |
| Culture | G37 | Exercise Patience | PS5, PS7, PS12, PS 13, PS17, PS 21, PS23, PS26, PS31, PS34, PS35, PS40, PS43, PS47, PS50, PS53, PS58, PS64 |
| | G38 | Educate executives at your company about the | PS7, PS16, PS 25, PS33, PS39, PS46, PS49, PS54, |

| | | benefits of DevOps, in order to gain resource and budget support | PS62 |
| | G39 | Cohesive team work to fill gap during Isolation changes | PS2, PS4, PS7, PS 14, PS 18, PS20, PS23, PS32, PS33, PS37, PS42, PS45, PS48, PS49, PS52, PS56, PS59, PS65, PS67 |
| | G40 | Keep All Teams on the Same Page | PS4, PS10, PS 18, PS27, PS31, PS39, PS42, PS49, PS51, PS56, PS63 |
| | G41 | Enterprises should focus on building a collaborative culture with shared goals | PS3, PS6, PS11, PS 17, PS19, PS 23, PS26, PS30, PS33, PS35, PS39, PS41, PS47, PS51, PS53, PS56, PS61, PS68 |
| | G42 | Consider DevOps to be a Cultural Change | PS3, PS9, PS 13, PS21, PS 27, PS33, PS37, PS42, PS45, PS48, PS50, PS56, PS64 |
| | G43 | Select DevOps "Champions" | PS1, PS2, PS11, PS 17, PS22, PS 28, PS34, PS37, PS40, PS42, PS48, PS49, PS55, PS60, PS62 |
| | G44 | Assess your organization's readiness to utilize a microservices architecture | PS2, PS4, PS10, PS 14, PS19, PS 24, PS25, PS30, PS36, PS39, PS43, PS45, PS49, PS54, PS57, PS69 |
| | G45 | Become a Psychologist | PS5, PS11, PS 16, PS24, PS 26, PS32, PS39, PS42, PS48, PS54, PS59, PS65 |
| | G46 | Commit daily, reduce branching | PS8, PS16, PS 26, PS34, PS 39, PS46, PS50, PS58, PS68 |
| | G47 | Understand and address your unique needs | PS7, PS14, PS20, PS 25, PS36, PS 44, PS46, PS51, PS56 |
| | G48 | Start toward Your Business Goals | PS10, PS19, PS 27, PS37, PS45, PS52, PS61, PS62, PS70 |

*4.1. Results of Empirical investigations*

*4.1.1 Respondents bibliographic information*

The important aspects of bibliographic data of the survey participates are analyzed to check authenticity and generalizability of the collected data. The detailed bibliographic data of survey participants is given in Appendix-C.

*Respondent's country affiliation*

According to the analyzed results of bibliographic data, the survey participant is from 20 different countries. We noted that most respondents are from Asian countries. Besides, a good mix of the survey participates observed from over the globe (Figure 7), which is a positive sign for the generalization of study results. Based on the respondent's country affiliation, we are confident as the results of our study can consider by software industry of any country.

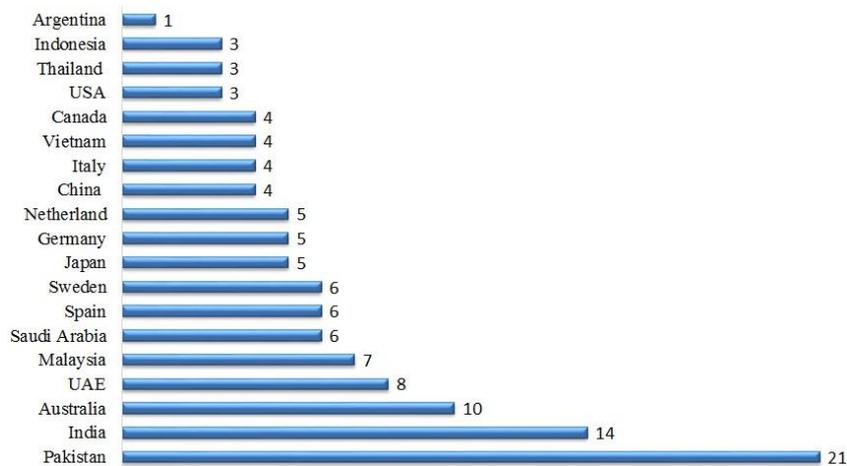

Figure 7: Respondent's affiliation countries

*Respondent's organization size*

We further analyzed the bibliographic data to check the organization size of survey participants. The results presented in Figure 8 shows that 26 (22%) respondents belong to small organizations, 49 (42%) are belongs to medium organizations, and 41 (35%) are from large scale organizations. The results show that there is a significant proportion of survey participants form each size of the organization. Hence, it is concluded that the results of this study are useful for every size of the organization.

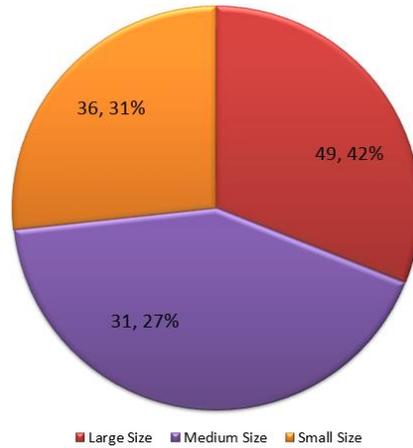
Figure 8: Respondents organization size

### *Respondents working experience*
The bibliographic data was also investigated to check the experiences of survey respondents. Figure 9 depicts the survey respondents' experience, which runs from two to twenty years. The statistics (6 and 5.5, respectively) represent a young group of people who responded. According to the researcher, "there is a good mix of survey participants with varied levels of experience with software development operations."

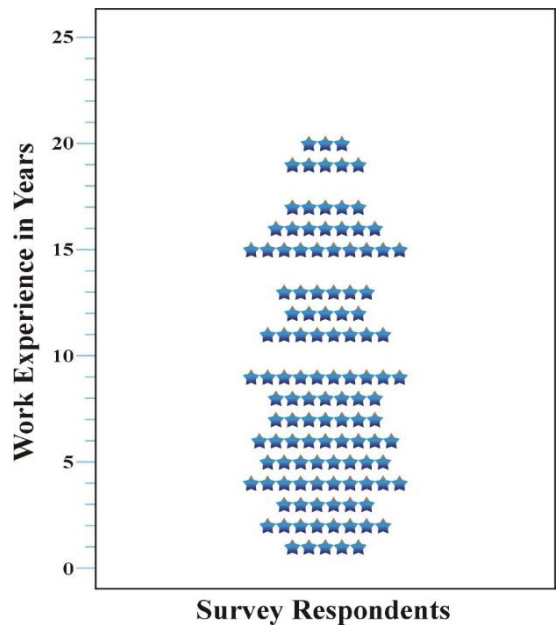
Figure 9: Experience of survey respondents

### *Respondent's designations*
Finstad et al. [38] mention that the responses are varied with respect to the designation of participants. Niazi et al. [31] reported that a respondent could only be measured appropriately if the participants deal with it frequently. The analyzed results show that most of the survey participants either project manager or software developers. The detailed results are shown in Figure 10.

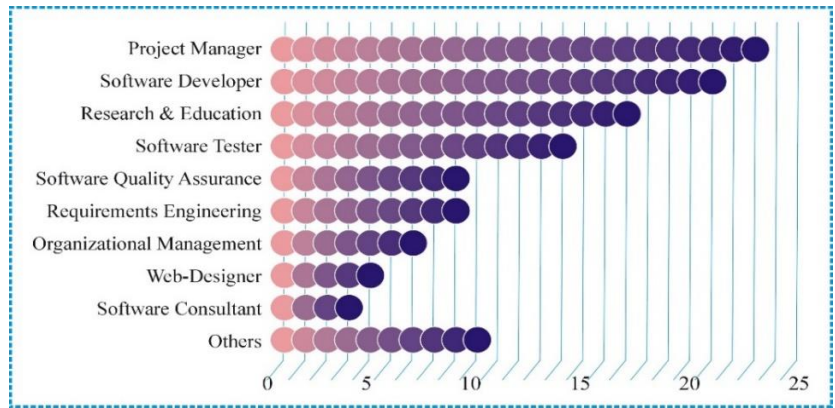
Figure 10: Designation of survey participants

### 4.1.2 Respondents feedback

The aim of this questionnaire survey was to collect expert input and perceptions on the selected guidelines and their related principles. A total of 116 full responses were considered for further analysis during the data gathering procedure. Positive (strongly agree, agree), Neutral (strongly disagree, disagree), and Negative (strongly disagree, disagree) responses were categorized (Table 5). The positive category findings indicate the ideas of those participants who agree with the identified criteria identified by SLR and their categories. The negative comments reflect the opinions of those who disagree with the established rules in the various categories. The results of the neutral category show the responses of those participants who have no idea regarding the impact of the mentioned factors.

The results of the empirical study presented in Table 5 shows that most of the survey participants agree as the reported guidelines could positively influence for sustainable DevOps implementation in in software organizations. It is observed that G41 (enterprises should focus on building a collaborative culture with shared goals, 91%) is reported as the most important guidelines from the survey participants. We further noted that G9 (Emphasize Quality Assurance Early, 88%) and G40 (Keep All Teams on the Same Page, 88%) are declared as the second highly considered guidelines by the survey respondents.

Moreover, it is noted that C4 (Culture, 93%) is the most important category of the investigated guidelines as considered by the survey participants. C3 (Sharing, 88%) and C1 (Measurement, 84%) is declared as the second and third highest regarded as principle of the guidelines considered by the survey participants.

Table 5: Results of a questionnaire survey study

|  | Number of responses =116 ||||||||
|  | Positive ||| Negative ||| Neutral ||
| S. No. | S. A | A | % | D | S. D | % | F | % |
| --- | --- | --- | --- | --- | --- | --- | --- | --- |
| C1 | 40 | 57 | 84 | 2 | 6 | 7 | 11 | 9 |
| G1 | 40 | 56 | 83 | 4 | 1 | 4 | 15 | 13 |
| G2 | 51 | 42 | 80 | 5 | 5 | 9 | 13 | 11 |
| G3 | 46 | 39 | 73 | 9 | 3 | 10 | 19 | 16 |
| G4 | 40 | 57 | 84 | 2 | 6 | 7 | 11 | 9 |
| G5 | 37 | 51 | 76 | 6 | 4 | 9 | 18 | 16 |
| G6 | 49 | 34 | 72 | 7 | 6 | 11 | 20 | 17 |
| G7 | 37 | 48 | 73 | 6 | 7 | 11 | 18 | 16 |
| G8 | 31 | 61 | 79 | 3 | 6 | 8 | 15 | 13 |
| G9 | 58 | 44 | 88 | 0 | 3 | 3 | 11 | 9 |
| G10 | 41 | 47 | 76 | 7 | 6 | 11 | 15 | 13 |
| G11 | 30 | 64 | 81 | 2 | 6 | 7 | 14 | 12 |
| C2 | 41 | 54 | 82 | 3 | 6 | 8 | 12 | 10 |
| G12 | 39 | 46 | 73 | 8 | 7 | 13 | 16 | 14 |
| G13 | 39 | 48 | 75 | 6 | 8 | 12 | 15 | 13 |
| G14 | 30 | 51 | 70 | 10 | 5 | 13 | 20 | 17 |
| G15 | 39 | 44 | 72 | 14 | 7 | 18 | 12 | 10 |
| G16 | 33 | 40 | 63 | 16 | 5 | 18 | 22 | 19 |
| G17 | 42 | 53 | 82 | 6 | 2 | 7 | 13 | 11 |
| G18 | 39 | 56 | 82 | 8 | 3 | 9 | 10 | 9 |
| G19 | 51 | 47 | 84 | 2 | 3 | 4 | 13 | 11 |
| G20 | 45 | 48 | 80 | 4 | 4 | 7 | 15 | 13 |
| G21 | 33 | 56 | 77 | 8 | 4 | 10 | 15 | 13 |
| G22 | 39 | 53 | 79 | 9 | 5 | 12 | 10 | 9 |
| G23 | 41 | 55 | 83 | 7 | 6 | 11 | 7 | 6 |
| G24 | 35 | 50 | 73 | 7 | 9 | 14 | 15 | 13 |
| G25 | 43 | 44 | 75 | 9 | 4 | 11 | 16 | 14 |
| G26 | 31 | 61 | 79 | 3 | 6 | 8 | 15 | 13 |
| C3 | 58 | 44 | 88 | 0 | 3 | 3 | 11 | 9 |
| G27 | 41 | 47 | 76 | 6 | 7 | 11 | 15 | 13 |
| G28 | 30 | 64 | 81 | 2 | 6 | 7 | 14 | 12 |
| G29 | 41 | 47 | 76 | 6 | 7 | 11 | 15 | 13 |
| G30 | 39 | 46 | 73 | 8 | 7 | 13 | 16 | 14 |
| G31 | 39 | 48 | 75 | 7 | 7 | 12 | 15 | 13 |
| G32 | 37 | 55 | 79 | 7 | 4 | 9 | 13 | 11 |
| G33 | 39 | 49 | 76 | 5 | 7 | 10 | 16 | 14 |
| G34 | 40 | 47 | 75 | 6 | 4 | 9 | 19 | 16 |
| G35 | 46 | 39 | 73 | 8 | 4 | 10 | 19 | 16 |
| G36 | 40 | 57 | 84 | 2 | 6 | 7 | 11 | 9 |
| C4 | 47 | 61 | 93 | 0 | 0 | - | 8 | 7 |
| G37 | 49 | 34 | 72 | 7 | 6 | 11 | 20 | 17 |
| G38 | 37 | 48 | 73 | 6 | 7 | 11 | 18 | 16 |
| G39 | 31 | 61 | 79 | 3 | 6 | 8 | 15 | 13 |
| G40 | 58 | 44 | 88 | 0 | 3 | 3 | 11 | 9 |
| G41 | 47 | 59 | 91 | 0 | 4 | 3 | 6 | 5 |
| G42 | 39 | 48 | 75 | 6 | 8 | 12 | 15 | 13 |
| G43 | 39 | 44 | 72 | 6 | 14 | 17 | 13 | 11 |
| G44 | 39 | 50 | 77 | 9 | 7 | 14 | 11 | 9 |
| G45 | 36 | 52 | 76 | 6 | 4 | 9 | 18 | 16 |
| G46 | 46 | 39 | 73 | 8 | 4 | 10 | 19 | 16 |
| G47 | 40 | 57 | 84 | 2 | 6 | 7 | 11 | 9 |
| G48 | 37 | 51 | 76 | 6 | 4 | 9 | 18 | 16 |

C1 =Measurement, C2= Automation, C3=Sharing, C4=Culture

### *4.2 Application of Fuzzy-AHP*
The fuzzy-AHP method was used to rank the discovered guidelines in terms of their importance in ensuring the sustainability of DevOps adopting in software industry. The phases of the Fuzzy-AHP technique are outlined in the sections below.

**Step-1 (Develop a hierarchy structure of reported guidelines and their categories)**

To apply the fuzzy-AHP, the critical session making problem is arranged in a hierarchy structure (as presented in Figure 5). The proposed hierarchy structure (Figure 11) was developed by considering the investigated guidelines and their core principle. The main objective of the study is found on the first levels (i.e., prioritization of sustainable DevOps guidelines), the categories and their corresponding guidelines are given on level-2 and level-3, respectively. The proposed hierarchy structure is presented in Figure 11.

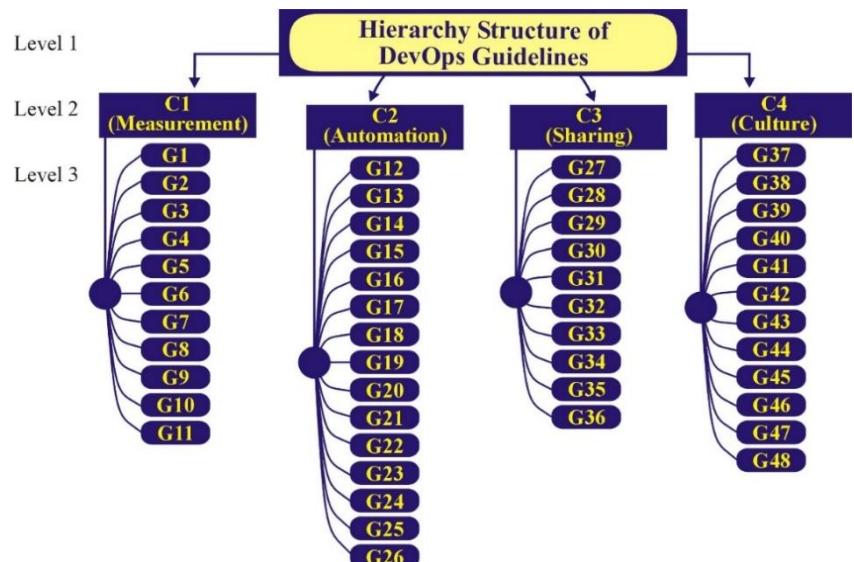

Figure 11. Proposed hierarchy structure

**Step-2 (Conducting the pairwise comparison)**

The motive of this study is to rank the identified guidelines in terms of their importance for long-term DevOps implementation in software development organizations. We created a questionnaire and contacted respondents from the initial survey to undertake the pairwise comparison (for fuzzy-AHP analysis). Participants in the survey provided a total of 29 replies. To ensure that no data was missing, all the responses were manually examined. All 29 replies were judged to be complete. Appendix-C includes instrument used for the pairwise-comparison (second survey).

With the application of fuzzy-AHP analysis, a small sample size can be a problem. However, several other studies [41-44] have used a dataset of similar size to do the AHP analysis. Based on the opinions of five experts, Shameem et al. [33] prioritized the agile software development influencing factors.

Cheng and Li [43] prioritize the success criteria of construction collaboration based on information gathered from nine experts. Lam and Zhao [44] collected feedback from eight experts to prioritize the influence teaching quality factors. Furthermore, Wong and Li [63] used an AHP analysis to pick an intelligent buildings system, considering the responses of nine experts. As a result, we conducted a fuzzy AHP analysis using data from 29 experts, which is a sufficient sample size for generalizing the findings of this study.

To analyze the pairwise comparison of the DevOps standards and their respective categories, the data acquired via the fuzzy AHP survey was transformed in geometric mean. The geometric mean is a valuable tool for converting expert opinions into TFN values; the formula for doing so is as follows:

$$GM = \sqrt[n]{m1 \times m2 \times m3 \ldots \ldots mn} \quad (17)$$
Where, m=response, n=responses

Table 6 shows the linguistic variables in relation to their triangular fuzzy Likert scales. The triangular fuzzy conversion scale (Table 6) proposed by Bozbura et al. [64] was used to build the pairwise comparison matrixes of the reported guidelines and their respective concept."

**Step-3 (Calculating the local priority weight of each guideline and their respective principle: A numerical example)**

The priority vector is calculated using the pairwise comparison matrix. The pairwise comparisons of the guidelines' principle are presented in Table 6 and the priority vector of the principles of guidelines presented in Table 9. Local Priority Weight (LPW) of all the principle of the guidelines were calculated using Equation 3. First, the synthetic extent values of four principle, i.e., measurement, automation, sharing and culture, were calculated and using equation 4, the priority is determined, and the used steps are given below. For TFN conversion scale, we used the guidelines of Bozbura [64]

$$S = (4, 5.1, 6.5) \otimes (0.04386, 0.054945, 0.070922) = (0.175439, 0.280220, 0.460993)$$

$$M = (2.2, 2.5, 3.2) \otimes (0.04386, 0.054945, 0.070922) = (0.096491, 0.137363, 0.226950)$$

$$C = (2.9, 3.6, 4.6) \otimes (0.04386, 0.054945, 0.070922) = (0.127193, 0.197802, 0.326241)$$

$$\sum_{i}^{n}\sum_{j}^{m} F_{gi}^{j} = (1,1,1) + (1.5, 2, 2.5) + (1, 1.5, 2)\ldots + (0.5, 0.6, 1) + (1, 1, 1) = (14.1, 18.2, 22.8)$$

$$\left[\sum_{i}^{n}\sum_{j}^{m} F_{gi}^{j}\right]^{-1} = (\frac{1}{22.8}, \frac{1}{18.2}, \frac{1}{14.1}) = (0.04386, 0.054945, 0.070922)$$

$$\sum_{j=1}^{m} F_{g1}^{j} = (1,1,1) + (1.5, 2.5, 3) + (1, 1.5, 2) + (1.5, 2.0, 2.5) = (5, 7, 8.5)$$

$$\sum_{j=1}^{m} F_{g2}^{j} = (0.3, 0.4, 0.6) + (1,1,1) + (0.4, 0.5, 0.6) + (0.5, 0.6, 1) = (2.2, 2.5, 3.2)$$

$$\sum_{j=1}^{m} F_{g3}^{j} = (0.5, 0.6, 1) + (1.5, 2, 2.5) + (1,1,1) + (1, 1.5, 2) = (4, 5.1, 6.5)$$

$$\sum_{j=1}^{m} F_{g4}^{j} = (0.4, 0.5, 0.6) + (1, 1.5, 2) + (0.5, 0.6, 1) + (1,1,1) = (2.9, 3.6, 4.6)$$

The "Measurement" (M), "Automation" (A), "Sharing" (S) and "Culture" (C) represent the synthesis values of DevOps principles which were calculated using Equation 4 as follow:

$$A = \sum_{j}^{m} F_{g1}^{j} \otimes \left[ \sum_{i}^{n} \sum_{j}^{m} F_{gi}^{j} \right]^{-1}$$

$$= (5, 7, 8.5) \otimes (0.04386, 0.054945, 0.070922) = (0.219298, 0.384615, 0.602837)$$

Table 6: Pairwise comparing between the principles

|  | Measurement | Automation | Sharing | Culture |
|---|---|---|---|---|
| Measurement | (1,1,1) | (0.3, 0.4, 0.5) | (1, 1.5, 2) | (0.5, 0.6, 0.1) |
| Automation | (2, 2.5, 3) | (1,1,1) | (0.4, 0.5, 0.6) | (0.5, 0.6, 0.1) |
| Sharing | (0.5, 0.6, 0.1) | (1.5, 2, 2.5) | (1,1,1) | (0.5, 0.6, 0.1) |
| Culture | (1, 1.5, 2) | (1, 1.5, 2) | (1, 1.5, 2) | (1,1,1) |

The level of possibility was calculated via equation 6, and equation 7 and 8 are used to calculate the priority-weight of pairwise comparison matrixes.

Table 7: Results of V values for criteria.

|  | M | A | S | C | d (Priority Weight) |
|---|---|---|---|---|---|
| V (M≥….) | - | 1 | 1 | 1 | 1 |
| V (A≥….) | 0.030018 | - | 0.26503 | 0.62273 | 0.030018 |
| V (S≥….) | 0.69837 | 1 | - | 1 | 0.69837 |
| V (C ≥….) | 0.36406 | 1 | 0.64662 | - | 0.36406 |

The calculated weight vector is W' = (1, 0.030019, 0.69836, 0.36405) (Table 7) and by normalizing weight vector, priority wight are determined i.e., W = (0.4789, 0.01435, 0.3337). As per calculated weights, "culture" is the highest priority principle of selected DevOps project management guidelines.

**Step-4 (Test the consistency of the pairwise matrix)**

All pairwise comparison matrices are evaluated for consistency, and we offered a step-by-step calculation of the technique used to determine whether a particular pairwise matrix is consistent. We considered the Table of Principles for this i.e., (Table 8). Using Equation 14, a triangular fuzzy number from the pairwise comparison matrix of the DevOps principles is defuzzified to a crisp number, corresponding the Fuzzy Crisp Matrix (FCM), as shown in Table 8:

Table 8: FCM for DevOps principle

|  | Measurement | Automation | Sharing | Culture |
|---|---|---|---|---|
| Measurement | 1.0 | 2.5 | 1.5 | 2.0 |
| Automation | 0.5 | 1.0 | 0.5 | 0.7 |
| Sharing | 0.7 | 2.0 | 1.0 | 1.5 |
| Culture | 0.5 | 1.5 | 0.7 | 1.0 |
| Column Sum | 2.7 | 7.0 | 3.7 | 5.2 |

The value of $\lambda_{max}$ of FCM matrix is calculated by adding the values of each column of Table 8, and then divided each value of with its column sum. At the end, the average of each is calculated for priority weight calculation (Table 9).

Table 9: Normalized matrix of DevOps guidelines

|  | Measurement | Automation | Sharing | Culture | Priority |
|---|---|---|---|---|---|
| Measurement | (1,1,1) | (0.3, 0.4, 0.5) | (1, 1.5, 2) | (0.5, 0.6, 0.1) | 0.11591 |
| Automation | (2, 2.5, 3) | (1,1,1) | (0.4, 0.5, 0.6) | (0.5, 0.6, 0.1) | 0.29500 |
| Sharing | (0.5, 0.6, 0.1) | (1.5, 2, 2.5) | (1,1,1) | (0.5, 0.6, 0.1) | 0.17028 |

| Culture | (1, 1.5, 2) | (1, 1.5, 2) | (1, 1.5, 2) | (1,1,1) | 0.41882 |

$$\lambda_{max}= \Sigma\ ([\Sigma\ Cj] \times \{W\}) \qquad (18)$$

Where, ΣCj= sum of the columns of Matrix [C] (Table 6),

W= weight-vector (Table 9), therefore

$\lambda_{max}$ = 2.7*0.11591+ 7.0*0.29500+ 3.7*0.17028+ 5.2*0.41882= 4.1067

According to the calculations, the FCM maximum Eigenvalue (max) is 4.1067. The FCM has 4 number. As a result, n=4 and the Random Consistency Index (RI) for n=4 is 0.9. (Table 3). As a result, the consistency index and consistency ration are calculated using equations 15 and 16 as follows:

$$CI = \frac{\lambda_{max} - n}{n-1} = \frac{4.1067 - 4}{4-1} = 0.035553$$

$$CR = \frac{CI}{RI} = \frac{0.035553}{0.9} = 0.039503$$

The determined CR is 0.039513<0.10; therefore, the developed pairwise matrixes are constant. Using the same process, the consistency of other matrixes is determined and given at the end of Table 10, 11,12 and 13.

Table 10: Pairwise comparison of guidelines of Measurement principle

|    | G1 | G2 | G3 | G4 | G5 | G6 | G7 | G8 | G9 | G10 | G11 | Priority |
|----|----|----|----|----|----|----|----|----|----|-----|-----|----------|
| G1 | (1,1,1) | (1, 1.5, 2) | (0.4, 0.5, 0.6) | (1, 1.5, 2) | (0.5, 0.6, 1) | (1.5, 2, 2.5) | (1, 1.5, 2) | (0.5, 0.6, 1) | (1.5, 2, 2.5) | (1, 1.5, 2) | (0.5, 0.6, 1) | 0.099531 |
| G2 | (0.5, 0.6, 1) | (1,1,1) | (1.5, 2, 2.5) | (0.5, 0.6, 1) | (0.4, 0.5, 0.6) | (1, 1.5, 2) | (1.5, 2, 2.5) | (0.5, 0.6, 1) | (1, 1.5, 2) | (0.4, 0.5, 0.6) | (1.5, 2, 2.5) | 0.095757 |
| G3 | (1.5, 2, 2.5) | (0.4, 0.5, 0.6) | (1,1,1) | (0.4, 0.5, 0.6) | (0.5, 0.6, 1) | (1.5, 2, 2.5) | (0.5, 0.6, 1) | (1.5, 2, 2.5) | (0.5, 0.6, 1) | (1, 1.5, 2) | (0.5, 0.6, 1) | 0.089031 |
| G4 | (0.5, 0.6, 1) | (1, 1.5, 2) | (0.5, 0.6, 1) | (1,1,1) | (1, 1.5, 2) | (0.5, 0.6, 1) | (0.5, 0.6, 1) | (1, 1.5, 2) | (0.5, 0.6, 1) | (2, 2.5, 3) | (1, 1.5, 2) | 0.094217 |
| G5 | (1, 1.5, 2) | (1.5, 2, 2.5) | (1, 1.5, 2) | (0.5, 0.6, 1) | (1,1,1) | (0.4, 0.5, 0.6) | (1.5, 2, 2.5) | (1.5, 2, 2.5) | (0.5, 0.6, 1) | (0.5, 0.6, 1) | (1.5, 2, 2.5) | 0.106180 |
| G6 | (0.4, 0.5, 0.6) | (0.5, 0.6, 1) | (0.4, 0.5, 0.6) | (1, 1.5, 2) | (1, 1.5, 2) | (1,1,1) | (0.4, 0.5, 0.6) | (1, 1.5, 2) | (0.4, 0.5, 0.6) | (1, 1.5, 2) | (0.5, 0.6, 1) | 0.073984 |
| G7 | (0.5, 0.6, 1) | (0.4, 0.5, 0.6) | (1, 1.5, 2) | (1, 1.5, 2) | (0.5, 0.6, 1) | (1.5, 2, 2.5) | (1,1,1) | (0.4, 0.5, 0.6) | (0.5, 0.6, 1) | (1.5, 2, 2.5) | (0.5, 0.6, 1) | 0.085232 |
| G8 | (1, 1.5, 2) | (1, 1.5, 2) | (0.4, 0.5, 0.6) | (0.5, 0.6, 1) | (0.4, 0.5, 0.6) | (0.5, 0.6, 1) | (1.5, 2, 2.5) | (1,1,1) | (1.5, 2, 2.5) | (1, 1.5, 2) | (1, 1.5, 2) | 0.098665 |
| G9 | (0.4, 0.5, 0.6) | (0.4, 0.5, 0.6) | (1, 1.5, 2) | (1, 1.5, 2) | (1, 1.5, 2) | (1.5, 2, 2.5) | (1, 1.5, 2) | (0.4, 0.5, 0.6) | (1,1,1) | (0.4, 0.5, 0.6) | (0.5, 0.6, 1) | 0.085852 |
| G10 | (0.5, 0.6, 1) | (1.5, 2, 2.5) | (0.5, 0.6, 1) | (0.3, 0.4, 0.5) | (1, 1.5, 2) | (0.5, 0.6, 1) | (0.4, 0.5, 0.6) | (0.5, 0.6, 1) | (1.5, 2, 2.5) | (1,1,1) | (1.5, 2, 2.5) | 0.088277 |
| G11 | (1, 1.5, 2) | (0.4, 0.5, 0.6) | (1, 1.5, 2) | (0.5, 0.6, 1) | (0.4, 0.5, 0.6) | (1, 1.5, 2) | (1, 1.5, 2) | (0.5, 0.6, 1) | (1, 1.5, 2) | (0.4, 0.5, 0.6) | (1,1,1) | 0.083275 |

λ= 12.249, CI = 0.12485, CR = 0.082685

Table 11: Pairwise comparison of guidelines of Automation principle

|     | G12 | G13 | G14 | G15 | G16 | G17 | G18 | G19 | G20 | G21 | B22 | G23 | G24 | G25 | G26 | Priority |
|-----|-----|-----|-----|-----|-----|-----|-----|-----|-----|-----|-----|-----|-----|-----|-----|----------|
| G12 | (1,1,1) | (0.5, 0.6, 1) | (0.4, 0.5, 0.6) | (1.5, 2, 2.5) | (0.4, 0.5, 0.6) | (0.4, 0.5, 0.6) | (1.5, 2, 2.5) | (1.5, 2, 2.5) | (0.4, 0.5, 0.6) | (1.5, 2, 2.5) | (2.5, 3, 3.5) | (1, 1.5, 2) | (1.5, 2, 2.5) | (0.4, 0.5, 0.6) | (1, 1.5, 2) | 0.078232 |
| G13 | (1, 1.5, 2) | (1,1,1) | (1.5, 2, 2.5) | (0.5, 0.6, 1) | (0.4, 0.5, 0.6) | (1.5, 2, 2.5) | (0.5, 0.6, 1) | (1.5, 2, 2.5) | (0.5, 0.6, 1) | (2.5, 3, 3.5) | (1.5, 2, 2.5) | (0.4, 0.5, 0.6) | (1, 1.5, 2) | (0.4, 0.5, 0.6) | (1, 1.5, 2) | 0.077156 |
| G14 | (1, 1.5, 2) | (0.4, 0.5, 0.6) | (1,1,1) | (0.4, 0.5, 0.6) | (1, 1.5, 2) | (0.4, 0.5, 0.6) | (1, 1.5, 2) | (0.2, 0.3, 0.4) | (0.4, 0.5, 0.6) | (1, 1.5, 2) | (1.5, 2, 2.5) | (0.5, 0.6, 1) | (1.5, 2, 2.5) | (0.4, 0.5, 0.6) | (1, 1.5, 2) | 0.061135 |
| G15 | (0.4, 0.5, 0.6) | (1, 1.5, 2) | (1.5, 2, 2.5) | (1,1,1) | (1, 1.5, 2) | (0.4, 0.5, 0.6) | (1.5, 2, 2.5) | (0.5, 0.6, 1) | (1.5, 2, 2.5) | (1, 1.5, 2) | (0.5, 0.6, 1) | (1.5, 2, 2.5) | (1, 1.5, 2) | (0.5, 0.6, 1) | (0.5, 0.6, 1) | 0.072116 |
| G16 | (1.5, 2, 2.5) | (1.5, 2, 2.5) | (0.4, 0.5, 0.6) | (0.5, 0.6, 1) | (1,1,1) | (1.5, 2, 2.5) | (0.5, 0.6, 1) | (0.4, 0.5, 0.6) | (1, 1.5, 2) | (1.5, 2, 2.5) | (0.5, 0.6, 1) | (1, 1.5, 2) | (0.5, 0.6, 1) | (1.5, 2, 2.5) | (1.5, 2, 2.5) | 0.075751 |
| G17 | (1.5, 2, 2.5) | (0.4, 0.5, 0.6) | (1.5, 2, 2.5) | (1.5, 2, 2.5) | (0.4, 0.5, 0.6) | (1,1,1) | (1, 1.5, 2) | (0.4, 0.5, 0.6) | (1, 1.5, 2) | (0.5, 0.6, 1) | (1, 1.5, 2) | (1, 1.5, 2) | (0.5, 0.6, 1) | (1.5, 2, 2.5) | (1, 1.5, 2) | 0.074993 |
| G18 | (0.4, 0.5, 0.6) | (1, 1.5, 2) | (0.4, 0.5, 0.6) | (0.4, 0.5, 0.6) | (1, 1.5, 2) | (0.5, 0.6, 1) | (1,1,1) | (1, 1.5, 2) | (0.4, 0.5, 0.6) | (1.5, 2, 2.5) | (0.5, 0.6, 1) | (1.5, 2, 2.5) | (1, 1.5, 2) | (0.5, 0.6, 1) | (1.5, 2, 2.5) | 0.065516 |
| G19 | (0.4, 0.5, 0.6) | (0.4, 0.5, 0.6) | (2.5, 3, 3.5) | (1, 1.5, 2) | (1.5, 2, 2.5) | (1.5, 2, 2.5) | (0.5, 0.6, 1) | (1,1,1) | (1.5, 2, 2.5) | (0.5, 0.6, 1) | (0.4, 0.5, 0.6) | (1, 1.5, 2) | (1.5, 2, 2.5) | (0.5, 0.6, 1) | (1, 1.5, 2) | 0.077156 |
| G20 | (1.5, 2, 2.5) | (1, 1.5, 2) | (1.5, 2, 2.5) | (0.4, 0.5, 0.6) | (0.5, 0.6, 1) | (0.5, 0.6, 1) | (1.5, 2, 2.5) | (0.4, 0.5, 0.6) | (1,1,1) | (1, 1.5, 2) | (0.4, 0.5, 0.6) | (1, 1.5, 2) | (0.5, 0.6, 1) | (1, 1.5, 2) | (1, 1.5, 2) | 0.069726 |
| G21 | (0.4, 0.5, 0.6) | (0.2, 0.3, 0.4) | (0.4, 0.5, 0.6) | (0.5, 0.6, 1) | (0.4, 0.5, 0.6) | (1, 1.5, 2) | (0.4, 0.5, 0.6) | (1, 1.5, 2) | (0.5, 0.6, 1) | (1,1,1) | (1, 1.5, 2) | (0.4, 0.5, 0.6) | (1.5, 2, 2.5) | (0.5, 0.6, 1) | (1.5, 2, 2.5) | 0.052684 |
| G22 | (0.2, 0.3, 0.4) | (0.4, 0.5, 0.6) | (0.4, 0.5, 0.6) | (1, 1.5, 2) | (1, 1.5, 2) | (0.5, 0.6, 1) | (1, 1.5, 2) | (1.5, 2, 2.5) | (1.5, 2, 2.5) | (0.5, 0.6, 1) | (1,1,1) | (1, 1.5, 2) | (0.4, 0.5, 0.6) | (1, 1.5, 2) | (0.5, 0.6, 1) | 0.06264 |
| G23 | (0.5, 0.6, 1) | (1.5, 2, 2.5) | (1, 1.5, 2) | (0.4, 0.5, 0.6) | (0.4, 0.5, 0.6) | (0.5, 0.6, 1) | (0.4, 0.5, 0.6) | (0.5, 0.6, 1) | (0.5, 0.6, 1) | (1.5, 2, 2.5) | (0.5, 0.6, 1) | (1,1,1) | (1.5, 2, 2.5) | (0.4, 0.5, 0.6) | (0.4, 0.5, 0.6) | 0.052583 |

| | | | | | | | | | | | | | | | |
|---|---|---|---|---|---|---|---|---|---|---|---|---|---|---|---|
| G24 | (0.4, 0.5, 0.6) | (0.5, 0.6, 1) | (0.4, 0.5, 0.6) | (0.5, 0.6, 1) | (1, 1.5, 2) | (1, 1.5, 2) | (0.5, 0.6, 1) | (0.4, 0.5, 0.6) | (1, 1.5, 2) | (0.4, 0.5, 0.6) | (1.5, 2, 2.5) | (0.4, 0.5, 0.6) | (1,1,1) | (1.5, 2, 2.5) | (0.5, 0.6, 1) | 0.054597 |
| G25 | (1.5, 2, 2.5) | (1.5, 2, 2.5) | (1.5, 2, 2.5) | (1, 1.5, 2) | (0.4, 0.5, 0.6) | (0.4, 0.5, 0.6) | (1, 1.5, 2) | (1, 1.5, 2) | (0.5, 0.6, 1) | (1, 1.5, 2) | (0.5, 0.6, 1) | (1.5, 2, 2.5) | (0.4, 0.5, 0.6) | (1,1,1) | (0.4, 0.5, 0.6) | 0.071115 |
| G26 | (0.5, 0.6, 1) | (0.5, 0.6, 1) | (0.4, 0.5, 0.6) | (1, 1.5, 2) | (0.4, 0.5, 0.6) | (0.5, 0.6, 1) | (0.4, 0.5, 0.6) | (0.4, 0.5, 0.6) | (0.5, 0.6, 1) | (0.4, 0.5, 0.6) | (1, 1.5, 2) | (1.5, 2, 2.5) | (1, 1.5, 2) | (1.5, 2, 2.5) | (1,1,1) | 0.054597 |

$\lambda_{max}$ = 17.140, CI = 0.15286, CR = 0.09613

Table 12: Pairwise comparison of guidelines of Sharing principle

| | G27 | G28 | G29 | G30 | G31 | G32 | G33 | G34 | G35 | G36 | Priority |
|---|---|---|---|---|---|---|---|---|---|---|---|
| G27 | (1,1,1) | (1, 1.5, 2) | (0.4, 0.5, 0.6) | (1, 1.5, 2) | (0.5, 0.6, 1) | (1, 1.5, 2) | (1, 1.5, 2) | (0.5, 0.6, 1) | (1.5, 2, 2.5) | (1, 1.5, 2) | 0.110115 |
| G28 | (0.5, 0.6, 1) | (1,1,1) | (1.5, 2, 2.5) | (0.5, 0.6, 1) | (0.4, 0.5, 0.6) | (1, 1.5, 2) | (1.5, 2, 2.5) | (0.5, 0.6, 1) | (1, 1.5, 2) | (0.5, 0.6, 1) | 0.098379 |
| G29 | (1.5, 2, 2.5) | (0.4, 0.5, 0.6) | (1,1,1) | (0.4, 0.5, 0.6) | (0.5, 0.6, 1) | (1, 1.5, 2) | (0.4, 0.5, 0.6) | (1.5, 2, 2.5) | (0.5, 0.6, 1) | (1, 1.5, 2) | 0.095144 |
| G30 | (0.5, 0.6, 1) | (1, 1.5, 2) | (0.5, 0.6, 1) | (1,1,1) | (1.5, 2, 2.5) | (0.5, 0.6, 1) | (0.4, 0.5, 0.6) | (1.5, 2, 2.5) | (0.5, 0.6, 1) | (0.5, 0.6, 1) | 0.089664 |
| G31 | (1, 1.5, 2) | (1.5, 2, 2.5) | (1, 1.5, 2) | (0.4, 0.5, 0.6) | (1,1,1) | (0.4, 0.5, 0.6) | (1.5, 2, 2.5) | (1.5, 2, 2.5) | (0.5, 0.6, 1) | (0.5, 0.6, 1) | 0.109770 |
| G32 | (0.5, 0.6, 1) | (0.5, 0.6, 1) | (0.5, 0.6, 1) | (1, 1.5, 2) | (1, 1.5, 2) | (1,1,1) | (0.4, 0.5, 0.6) | (1, 1.5, 2) | (0.4, 0.5, 0.6) | (1, 1.5, 2) | 0.087082 |
| G33 | (0.5, 0.6, 1) | (0.4, 0.5, 0.6) | (1.5, 2, 2.5) | (1.5, 2, 2.5) | (1.5, 2, 2.5) | (1.5, 2, 2.5) | (1,1,1) | (0.5, 0.6, 1) | (0.5, 0.6, 1) | (1.5, 2, 2.5) | 0.119216 |
| G34 | (1, 1.5, 2) | (1, 1.5, 2) | (0.4, 0.5, 0.6) | (0.4, 0.5, 0.6) | (0.4, 0.5, 0.6) | (0.5, 0.6, 1) | (1, 1.5, 2) | (1,1,1) | (1.5, 2, 2.5) | (1, 1.5, 2) | 0.099249 |
| G35 | (0.4, 0.5, 0.6) | (0.4, 0.5, 0.6) | (1, 1.5, 2) | (1, 1.5, 2) | (1, 1.5, 2) | (1.5, 2, 2.5) | (1, 1.5, 2) | (0.4, 0.5, 0.6) | (1,1,1) | (0.4, 0.5, 0.6) | 0.097639 |
| G36 | (0.5, 0.6, 1) | (1, 1.5, 2) | (0.5, 0.6, 1) | (1, 1.5, 2) | (1, 1.5, 2) | (0.5, 0.6, 1) | (0.4, 0.5, 0.6) | (0.5, 0.6, 1) | (1.5, 2, 2.5) | (1,1,1) | 0.093742 |

$\lambda$ = 11.278, CI = 0.14197, CR = 0.095285

Table 13: Pairwise comparison of guidelines of Culture principle

| | G37 | G38 | G39 | G40 | G41 | G42 | G43 | G44 | G45 | G46 | G47 | G48 | Priority |
|---|---|---|---|---|---|---|---|---|---|---|---|---|---|
| G37 | (1,1,1) | (1, 1.5, 2) | (0.4, 0.5, 0.6) | (1.5, 2, 2.5) | (0.5, 0.6, 1) | (1.5, 2, 2.5) | (1, 1.5, 2) | (0.5, 0.6, 1) | (1.5, 2, 2.5) | (1, 1.5, 2) | (0.5, 0.6, 1) | (0.5, 0.6, 1) | 0.089771 |
| G38 | (0.5, 0.6, 1) | (1,1,1) | (1.5, 2, 2.5) | (0.5, 0.6, 1) | (0.4, 0.5, 0.6) | (1, 1.5, 2) | (1.5, 2, 2.5) | (0.5, 0.6, 1) | (1, 1.5, 2) | (0.5, 0.6, 1) | (1.5, 2, 2.5) | (1.5, 2, 2.5) | 0.092637 |
| G39 | (1.5, 2, 2.5) | (0.4, 0.5, 0.6) | (1,1,1) | (0.4, 0.5, 0.6) | (0.5, 0.6, 1) | (1.5, 2, 2.5) | (0.4, 0.5, 0.6) | (1.5, 2, 2.5) | (0.5, 0.6, 1) | (1, 1.5, 2) | (0.5, 0.6, 1) | (1, 1.5, 2) | 0.082375 |
| G40 | (0.4, 0.5, 0.6) | (1, 1.5, 2) | (0.5, 0.6, 1) | (1,1,1) | (1.5, 2, 2.5) | (0.5, 0.6, 1) | (0.5, 0.6, 1) | (1.5, 2, 2.5) | (0.5, 0.6, 1) | (0.4, 0.5, 0.6) | (1, 1.5, 2) | (0.5, 0.6, 1) | 0.074275 |
| G41 | (1, 1.5, 2) | (1.5, 2, 2.5) | (1, 1.5, 2) | (0.4, 0.5, 0.6) | (1,1,1) | (0.4, 0.5, 0.6) | (1.5, 2, 2.5) | (1.5, 2, 2.5) | (0.5, 0.6, 1) | (0.4, 0.5, 0.6) | (1.5, 2, 2.5) | (1.5, 2, 2.5) | 0.099306 |
| G42 | (0.4, 0.5, 0.6) | (0.5, 0.6, 1) | (0.4, 0.5, 0.6) | (1, 1.5, 2) | (1, 1.5, 2) | (1,1,1) | (0.5, 0.6, 1) | (1, 1.5, 2) | (0.4, 0.5, 0.6) | (1, 1.5, 2) | (0.5, 0.6, 1) | (1, 1.5, 2) | 0.072511 |
| G43 | (0.5, 0.6, 1) | (0.4, 0.5, 0.6) | (1.5, 2, 2.5) | (1, 1.5, 2) | (1.5, 2, 2.5) | (1, 1.5, 2) | (1,1,1) | (0.4, 0.5, 0.6) | (0.5, 0.6, 1) | (1.5, 2, 2.5) | (0.5, 0.6, 1) | (0.5, 0.6, 1) | 0.083396 |
| G44 | (1, 1.5, 2) | (1, 1.5, 2) | (0.4, 0.5, 0.6) | (0.4, 0.5, 0.6) | (0.4, 0.5, 0.6) | (0.5, 0.6, 1) | (1.5, 2, 2.5) | (1,1,1) | (1.5, 2, 2.5) | (1, 1.5, 2) | (1.5, 2, 2.5) | (1, 1.5, 2) | 0.093556 |
| G45 | (0.4, 0.5, 0.6) | (0.4, 0.5, 0.6) | (1, 1.5, 2) | (1, 1.5, 2) | (1, 1.5, 2) | (1.5, 2, 2.5) | (1, 1.5, 2) | (0.4, 0.5, 0.6) | (1,1,1) | (0.4, 0.5, 0.6) | (0.5, 0.6, 1) | (0.5, 0.6, 1) | 0.074588 |
| G46 | (0.5, 0.6, 1) | (1, 1.5, 2) | (0.5, 0.6, 1) | (1.5, 2, 2.5) | (1.5, 2, 2.5) | (0.5, 0.6, 1) | (0.4, 0.5, 0.6) | (0.5, 0.6, 1) | (1.5, 2, 2.5) | (1,1,1) | (1.5, 2, 2.5) | (1, 1.5, 2) | 0.092637 |
| G47 | (1, 1.5, 2) | (0.4, 0.5, 0.6) | (1, 1.5, 2) | (0.5, 0.6, 1) | (0.4, 0.5, 0.6) | (1, 1.5, 2) | (1, 1.5, 2) | (0.4, 0.5, 0.6) | (1, 1.5, 2) | (0.4, 0.5, 0.6) | (1,1,1) | (0.5, 0.6, 1) | 0.071128 |
| G48 | (1, 1.5, 2) | (0.4, 0.5, 0.6) | (0.5, 0.6, 1) | (1, 1.5, 2) | (0.4, 0.5, 0.6) | (0.5, 0.6, 1) | (1, 1.5, 2) | (0.5, 0.6, 1) | (1, 1.5, 2) | (0.5, 0.6, 1) | (1, 1.5, 2) | (1,1,1) | 0.073821 |

$\lambda$ = 13.542, CI = 0.14017, CR = 0.094709

**Step 5: Calculating the global weights**

The local weight (LW) represents the impact of a guideline on the overall study objective, i.e., prioritization of sustainable DevOps implementation guidelines, while the global weight (GW) represents the impact of a guideline on the overall study objective, i.e., prioritization of sustainable DevOps implementation guidelines. Beyond their principle, the GW is utilized to calculate the ultimate ranking of the guidelines in comparison to all 48 guidelines evaluated. The GW is calculated by multiplying a guideline's LW by the weight of their principal. For example, the LW of G1 (Organizations start DevOps practices with small projects, 0.099531) and the principal's weight is C1 (Measurement, 0.11591); so, the GW of G1 = (0.099531) × (0.11591) = 0.011537

(Table 14). By comparing the local rank of G1 within their mapped principle, it is ranked as the second-highest priority guideline.

While comparing its GW with all other 48 guidelines, it stands out 39$^{th}$ most important guideline for the sustainable DevOps implementation in software organizations. The results presented in Table 14 shows that the GW of G41 (Enterprises should focus on building a collaborative culture with shared goals, GW=0.041591) is the highest priority guideline for sustainable DevOps implementation in software organizations. Moreover, G44 (Assess your organization's readiness to utilize a microservices architecture, GW= 0.039183) and G 38 (Educate executives at your company about the benefits of DevOps, to gain resource and budget support, GW=0.038798) are declared as the second and third most significant guidelines for DevOps paradigm. The final ranking of all the other guidelines is presented in Table 14.

Table 14: Determining global weights

| Category | Principle Weight | Guidelines | Local Weight | Local Rank | Global Weight | Global Rank |
|---|---|---|---|---|---|---|
| C1 (Measurement) | 0.11591 | G1 | 0.099531 | 2 | 0.011537 | 39 |
| | | G2 | 0.095757 | 4 | 0.011099 | 41 |
| | | G3 | 0.089031 | 6 | 0.01032 | 43 |
| | | G4 | 0.094217 | 5 | 0.010921 | 42 |
| | | G5 | 0.106180 | 1 | 0.012307 | 38 |
| | | G6 | 0.073984 | 11 | 0.008575 | 48 |
| | | G7 | 0.085232 | 9 | 0.009879 | 46 |
| | | G8 | 0.098665 | 3 | 0.011436 | 40 |
| | | G9 | 0.085852 | 8 | 0.009951 | 45 |
| | | G10 | 0.088277 | 7 | 0.010232 | 44 |
| | | G11 | 0.083275 | 10 | 0.009652 | 47 |
| C2 (Automation) | 0.29500 | G12 | 0.078232 | 1 | 0.023078 | 13 |
| | | G13 | 0.077156 | 2 | 0.022761 | 14 |
| | | G14 | 0.061135 | 11 | 0.018035 | 26 |
| | | G15 | 0.072116 | 6 | 0.021274 | 18 |
| | | G16 | 0.075751 | 4 | 0.022347 | 16 |
| | | G17 | 0.074993 | 5 | 0.022123 | 17 |
| | | G18 | 0.065516 | 9 | 0.019327 | 22 |
| | | G19 | 0.077156 | 3 | 0.022761 | 15 |
| | | G20 | 0.069726 | 8 | 0.020569 | 20 |
| | | G21 | 0.052684 | 14 | 0.015542 | 34 |
| | | G22 | 0.06264 | 10 | 0.018479 | 25 |
| | | G23 | 0.052583 | 15 | 0.015512 | 35 |
| | | G24 | 0.054597 | 12 | 0.016106 | 31 |
| | | G25 | 0.071115 | 7 | 0.020979 | 19 |
| | | G26 | 0.054597 | 13 | 0.016106 | 32 |
| C3 (Sharing) | 0.17028 | G27 | 0.110115 | 2 | 0.01875 | 23 |
| | | G28 | 0.098379 | 5 | 0.016752 | 28 |
| | | G29 | 0.095144 | 7 | 0.016201 | 30 |
| | | G30 | 0.089664 | 9 | 0.015268 | 36 |
| | | G31 | 0.10977 | 3 | 0.018692 | 24 |
| | | G32 | 0.087082 | 10 | 0.014828 | 37 |
| | | G33 | 0.119216 | 1 | 0.0203 | 21 |
| | | G34 | 0.099249 | 4 | 0.0169 | 27 |
| | | G35 | 0.097639 | 6 | 0.016626 | 29 |
| | | G36 | 0.093742 | 8 | 0.015962 | 33 |
| C4 (Culture) | 0.41882 | G37 | 0.089771 | 5 | 0.037598 | 5 |
| | | G38 | 0.092637 | 3 | 0.038798 | 3 |
| | | G39 | 0.082375 | 7 | 0.0345 | 7 |
| | | G40 | 0.074275 | 9 | 0.031108 | 9 |
| | | G41 | 0.099306 | 1 | 0.041591 | 1 |
| | | G42 | 0.072511 | 11 | 0.030369 | 11 |
| | | G43 | 0.083396 | 6 | 0.034928 | 6 |
| | | G44 | 0.093556 | 2 | 0.039183 | 2 |
| | | G45 | 0.074588 | 8 | 0.031239 | 8 |
| | | G46 | 0.092637 | 4 | 0.038798 | 4 |
| | | G47 | 0.071128 | 12 | 0.02979 | 12 |
| | | G48 | 0.073821 | 10 | 0.030918 | 10 |

## 5. Summary and discussion

This study aims to explore the guidelines of sustainable DevOps implementation and to prioritize them concerning their significance for software organizations. To address the objective of this study, three research questions were proposed: the summary of the findings of each question is presented below:

***RQ1** (What guidelines for sustainable DevOps implementation in software development organizations are reported in the literature and industry practices?)*
A comprehensive literature review was done to identify prospective literature relevant to the study's goal. A total of 71 studies were found using the SLR. The studies were carefully reviewed, and 48 best practices that could influence the DevOps paradigm in the software industry were investigated. The guidelines were then classified into the CAMS model's (i.e., Culture, automation, measurement and sharing). The mapping of identified guidelines into CAMS aids in the development of the investigated guidelines' hierarchy structure, which is then used to perform the fuzzy-AHP. We conducted a questionnaire survey study with experts to scale the investigated SLR guidelines and their categorization in the CAMS model. 116 full responses were received from the experts during the data gathering process for the survey. All the researched guidelines from the current literature are related to industry practices, according to the summary results of the questionnaire survey study, respondents agree with the categories of the rules.

***RQ2** (How the explored guidelines were prioritized using fuzzy-AHP?)*
The fuzzy-AHP approaches' step-by-step protocols were used to prioritize investigated best practice in regard to their significance for the DevOps paradigm. The pairwise matrixes of the guidelines of each category were generated based on the expert's opinions to do the fuzzy-AHP analysis. All the fuzzy-AHP processes were meticulously followed, and the priority weights of each best practice were calculated. The priority weight (global weight) of each best practice was obtained using the fuzzy-AHP analysis. The results show that G41 (Enterprises should focus on building a collaborative culture with shared goals, GW=0.041591) is the highest priority best practice for DevOps adoption and its progression in software organizations. Leonardo et al.[6] highlighted that DevOps required a cultural change in the software development organization, as it offers continues and a collaborative work environment between the developers and operators. Gupta et al.[26] and Marijan et al.[65] also highlighted the importance of collaborative culture for the successful adoption of DevOps paradigm. Moreover, G44 (Assess your organization's readiness to utilize a microservices architecture, GW= 0.039183) and G38 (Educate executives at your company about the benefits of DevOps, to gain resource and budget support, GW=0.038798) are ranked as three most priority best practise of DevOps paradigm.

***RQ3** (What would be the prioritization-based framework of DevOps sustainable guidelines?)*
The taxonomy of the investigated guidelines was developed considering the CAMS model and using the rankings determined via fuzzy-AHP. We have used both global and local ranks (Table 14). The objective of taxonomy development is to show the impact of each guideline in their own principle and for overall DevOps paradigm.
For example, G1 (Organizations start DevOps practices with small projects) is locally ranked as the 2$^{nd}$ most important guideline for the sustainable DevOps execution in software industry. An interesting observation is that G1 is ranked as the 39$^{th}$. Similarly, G2 (Include modelling for legacy infrastructure and applications in your DevOps plans) is declared as the 4$^{th}$ most important guidelines in 'Management category' and ranked as the 42$^{nd}$ with respect to global rankings.
The local and global ranks of each guideline are presented in Figure 12, which renders the impact of a particular guidelines within their respective principal and for overall project compared with all the identified 48 guidelines. Moreover, C1 (Measurement, CW=0.41882) is ranked as the most significant principal of the identified guideline. Furthermore, it is observed that C2 (Automation=0.295), C3 (Sharing, CW=0.17028) is ranked as the second and third most significant principle of the sustainable DevOps implementation guidelines. Hence, the focus on these areas could assist the organization for sustainable DevOps practices implementations.

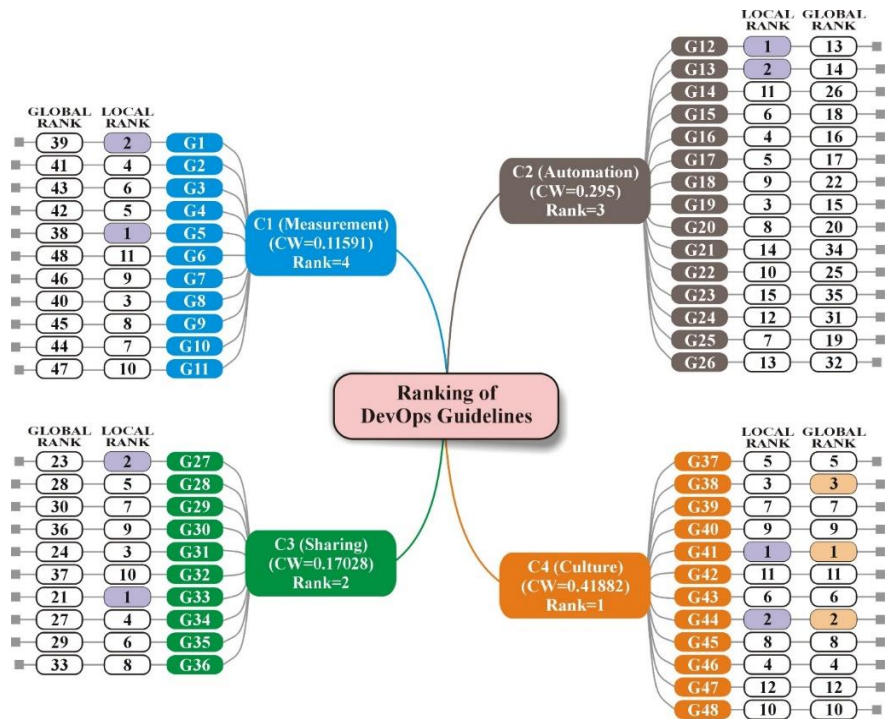
Figure 12: Prioritization based framework of the investigated guidelines

## 6. Threats to Validity

One of the limitations of the study is potential researcher's biasness in the investigated guidelines using a literature review study. To address this comment, the "inter-rater reliability test" was performed, and the results shows no significant biasness in the literature study findings. Moreover, the literature was collected from the limited selected repositories that might causes the chance of omitting the related literature. Though, considering the existing studies, this threat is no systematic [35, 40, 66].

Similarly, the small size of the data set poses an external threat to the generalization of the questionnaire survey. We received n=116 responses for this study, which may be insufficient to generalize the findings. However, based on previous research in the software engineering sector [17, 50, 51], this sample size is adequate for generalizing the findings.

Furthermore, the fuzzy-AHP was used to rank the investigated guidelines and their corresponding categories based on the opinions of experts. To mitigate this concern, the consistency ratio of pairwise comparison matrices was calculated, and the results show that fuzzy AHP analysis results have appropriate internal validity.

## 7. Conclusion and Future directions

DevOps is an approach which combines development and operations to enable agility during software development process. The implementation of DevOps practices is complex, and this motivates us to explore the guidelines that are important for sustainable DevOps implementation in software development organizations. A total of 48 guidelines were discovered as a consequence of the systematic literature review. The identified guidelines were then mapped into the CAMS model, which represents the key DevOps principles. Moreover, the questionnaire survey study was conducted to get the insight of experts on the identified guidelines. The results of the questionnaire survey study indicated that the identified guidelines are in line with real-world practices. Finally, the investigated guidelines were further prioritized with respect to their significance for sustainable DevOps implementation using fuzzy-AHP. The prioritization results show that 'enterprises should focus on building a collaborative culture with shared goals', 'assess your organization's readiness to utilize a microservices architecture' and 'educate executives at your company about the benefits of DevOps' are important guidelines. The categorization of investigated guidelines and their rankings provides a taxonomy that could assist the academic researchers and industry practitioners in revising and developing the new effective strategies for the sustainable implementation of DevOps paradigm in software organizations.

As part of future work, we plan to conduct multivocal literature review and case studies to explore the additional guidelines associated with DevOps paradigm. In addition, we will identify the critical challenges and success factors that need to be addressed for the sustainable DevOps practices in software organizations. Ultimately, we plan to develop a readiness model which will assist the practitioners in assessing and improving their DevOps implementation strategies.

**Appendixes:**
**Appendix-A:** Selected studies along with quality assessment score (https://tinyurl.com/y9x3fg3z)
**Appendix-B:** Sample of questionnaire survey (https://tinyurl.com/y832q5jy)
**Appendix-C:** Sample of pairwise comparison questionnaire (https://tinyurl.com/y97k7jp9)

**Conflict of Interest:** Authors declare no conflicts of interest.

**Acknowledgment**

The authors are grateful to the Deanship of Scientific Research, King Saud University for funding through Vice Deanship of Scientific Research Chairs.